\documentclass[showpacs,aps,graphicx]{revtex4}
\usepackage{graphicx}

\begin{document}

\title{Correlation dynamics of  a two-qubit system in a Bell-diagonal state under non-identical
local noises\footnote{published in Quantum Inf. Process.
\textbf{13}, 1175 -- 1189 (2014)}}

\author{Bao-Cang Ren, Hai-Rui Wei, and Fu-Guo Deng\footnote{Corresponding author: fgdeng@bnu.edu.cn}}
\address{ Department of Physics, Applied Optics Beijing Area Major
Laboratory, Beijing Normal University, Beijing 100875, China}

\date{\today }

\begin{abstract}
The property of quantum correlation has been studied in recent
years, especially for the quantum and classical correlations
affected by environment. The dynamics of quantum and classical
correlations in two-qubit system under identical local noise
channels have been investigated recently. Here we will consider the
dynamics of quantum and classical correlations when the local noise
channels of two sides are not identical. We investigate the dynamics
of quantum and classical correlations with three types of local
noise channels in both Markovian and non-Markovian conditions, and
show  the decay rules of quantum and classical correlations with
different types and parameter times of local noise
channels.\\

\end{abstract}
 \maketitle


\section*{  1 $\;\;$  Introduction}

Quantum entanglement is a kind of quantum correlation and it
 plays an important role in quantum information and
communication \cite{Nielsen}. However, entanglement is not the only
type of correlation useful for quantum technology, and there are
other nonclassical correlations apart from the entanglement, which
are also responsible for the quantum advantage over their classical
counterparts, such as deterministic quantum computation with one
pure qubit \cite{DQC1-1,DQC1-2,DQC1-3}, quantum phase transition
\cite{Phase} and quantum Grover research algorithm without
entanglement \cite{Research1,Research2}. Therefore, it is desirable
to investigate, characterize, and quantify quantum correlations.


It is widely accepted that mutual information is the total
correlation contained in a bipartite system, which is the sum of
classical correlation and quantum correlation
\cite{corr1,corr2,corr3,corr4}. Therefore, how to distinguish
between the quantum and the classical aspects of the total
correlation is an outstanding question. In 2001, based on the
distinction between the quantum information theory and classical
information theory, Olliver and Zurek \cite{Zurek} first defined a
quantifier, called the quantum discord, as a measure for all quantum
correlations   in bipartite systems. Other algorithms to evaluate
quantum correlations have also been investigated
\cite{pro1,pro2,pro3,pro4,GMQD1,GMQD2}. For pure states, quantum
discord  exactly equals to the entanglement, which implies that
there is no quantum correlation for separable pure states
\cite{corr1,corr2}. However, for general two-qubit mixed states, the
situation is more complicated. Up to now, quantum discord is also an
immature field. There are few analytical expressions, including
special cases for two-qubit system
\cite{analytical1,analytical2,analytical3}. For multiqubit and
high-dimension systems, it is still an open issue as how to define
quantum discord.

It is well known that a quantum system inevitably interacts with its
environment, which is responsible for the loss of quantum properties
initially  presenting  in  the  system.  That is, it is important to
investigate the behaviors of these quantum properties (quantum
entanglement and quantum correlations) under the action of
decoherence. For the entanglement dynamics of quantum systems, many
works have been done under the influence of both Markovian and
non-Markovian environments \cite{ent1,ent2,ent3,ent4,ent5}. It has
been shown that if the environment is a non-Markovian one,
entanglement can suddenly disappear at a finite time,  which was
named as "entanglement studden death" (ESD).  However, for a
Markovian one, ESD may  occur for a suitable choice of the noise
channel and initial states, but ESB can not be observed
\cite{ent4,ent5}. In recent years, quantum correlation dynamics has
received increasing attention, and similar works about the
entanglement dynamics have been investigated about the quantum
correlation
\cite{CorrDyna1,CorrDyna2,CorrDyna3,CorrDyna4,CorrDyna5,CorrDyna6,analytical3}.
The previous works showed  that quantum discord and entanglement
behave differently under the influence of the environment. In
particular, the phenomenon of ESD does not occur for quantum
discord, which disappears only asymptotically.


The dynamics of the discord in the presence of the environments
acting on each subsystems has, until now, been studied only for the
same type. In this paper, we consider the correlation dynamics of
two-qubit system under the effect of two non-identical independent
local Markovian and non-Markovian environments. When the Markovian
environments acting on two subsystems have the same type, there are
three regimes for different decay rules of quantum and classical
correlations with different relations of coefficients in initial
two-qubit state. For the case  with Markovian environments acting on
each subsystem belonging to different types, three are still three
coefficients regimes for decay rules of quantum and classical
correlations. However, if the environment is non-Markovian, there
are three coefficients regimes of decay rules for two subsystems
with the same type of local noise channels and only one coefficients
regime of decay rules for two subsystems with different types of
local noise channels. We focus on the coefficients regimes with
sudden change points on decay rates of quantum and classical
correlations, which are decided by the types and time parameters of
local noise channels.


This article is organized as follows. In
Sec.\uppercase\expandafter{\romannumeral2}, we introduce the
property of Bell-diagonal state and noise channels. The dynamics of
the quantum correlation under local Markovian environment will be
discussed in Sec.\uppercase\expandafter{\romannumeral3}. For
non-Markovian case, it is discussed in
Sec.\uppercase\expandafter{\romannumeral4}. Finally, we conclude our
work in Sec.\uppercase\expandafter{\romannumeral5}.


\section*{ 2 $\;\;$   Correlations and channels}

\subsection*{2.1  $\;\;$    Correlations}

The total information (mutual information) of a two-qubit bipartite
system is defined as
\begin{eqnarray}                               
I(\rho)=S(\rho_{A})+S(\rho_{B})-S(\rho),
\end{eqnarray}
where $\rho$ is the density matrix of the whole system $AB$, and
$\rho_A=\text{tr}_B\rho$ and $\rho_B=\text{tr}_A\rho$ are the
reduced density matrices of  the subsystems $A$ and $B$,
respectively.
\begin{eqnarray}                               
S(\rho)=-\text{tr}\rho\log_2\rho
\end{eqnarray}
is the von Neumann entropy for the matrix  $\rho$. Quantum discord
is defined as \cite{Zurek}
\begin{eqnarray}                                
Q(\rho)=I(\rho)-C(\rho),\label{discord}
\end{eqnarray}
where
\begin{eqnarray}                                
C(\rho)=\text{Max}_{\{B_k\}}(S(\rho_A)-S(\rho|\{B_k\}))\label{cc}
\end{eqnarray}
is the classical correlation of the state when acting the
measurements $\{B_k\}$ on the subsystem $B$.
\begin{eqnarray}                               
S(\rho|\{B_k\})=\sum_kp_kS(\rho_k)
\end{eqnarray}
is the conditional entropy of the subsystem $A$. Here
\begin{eqnarray}                               
\rho_k=\frac{(I\otimes B_k)\rho(I\otimes B_k)}{{tr(I\otimes
B_k)\rho(I\otimes B_k)}}
\end{eqnarray}
is the state of the system under the measurements $\{B_k\}$. Here
$I$ is the identity matrix.

In this work, we investigate the dynamics of two-qubit systems in a
Bell-diagonal state which has three parameters \cite{Caves}. It
includes subsets of separable states, classical states, and
entangled states with maximally mixed marginal
($\rho_{A(B)}=I_{A(B)}/2$), and it can be described as:
\begin{eqnarray}                                
\rho=\frac{1}{4}(I+\sum^3_{i=1}c_{i}\sigma_{i}\otimes\sigma_{i}),\label{eq7}
\end{eqnarray}
where $\sigma_i$ ($i=1,2,3$) are the three Pauli operators and $c_i$
represent the three parameters for describing Bell-diagonal states.
\begin{eqnarray}                               
\sigma_{1}= \left(\begin{array}{cc}
0&1\\
1&0\\
\end{array}
\right),\;\;\;\;\;\;\;\;\;\;\;\; \sigma_{2}= \left(\begin{array}{cc}
0&-i\\
i&0\\
\end{array}
\right),\;\;\;\;\;\;\;\;\;\;\;\; \sigma_{3}= \left(\begin{array}{cc}
1&0\\
0&-1\\
\end{array}
\right).
\end{eqnarray}
Here $i^2=-1$.  The state $\rho$ can also be written as
\begin{eqnarray}                               
\rho &= & \lambda_{\psi^+}|\psi^+\rangle\langle\psi^+| +
\lambda_{\psi^-}|\psi^-\rangle\langle\psi^-| 
+ \lambda_{\phi^+}|\phi^+\rangle\langle\phi^+| +
\lambda_{\phi^-}|\phi^-\rangle\langle\phi^-|,
\end{eqnarray}
where $|\psi^\pm\rangle$ and $|\phi^\pm\rangle$ are the four Bell
states for two-qubit systems, that is,
\begin{eqnarray}                               
|\psi^{\pm}\rangle &=& \frac{1}{\sqrt{2}}(|00\rangle \pm
|11\rangle), \nonumber\\
|\phi^{\pm}\rangle &=& \frac{1}{\sqrt{2}}(|01\rangle \pm
|10\rangle). 
\end{eqnarray}

The classical correlation and quantum correlation of these states
can be derived  from Eqs.(\ref{discord}) and (\ref{cc})
\cite{analytical3},
\begin{eqnarray}                                
C(\rho) &=&
\sum^2_{j=1}\frac{1+(-1)^j\chi}{2}\log_2[1+(-1)^j\chi],\label{eq11}\\
Q(\rho) &=&
2+\sum^4_{k=1}\lambda_k\log_2\lambda_k-C(\rho),\label{eq12}
\end{eqnarray}
where $\chi=\text{Max}\{|c_1|,|c_2|,|c_3|\}$. Classical correlation
and quantum correlation of this two-qubit system depend on the
coefficients $c_i$ $(i=1,2,3)$.

\subsection*{2.2  $\;\;$   Markovian  noise  channels}

We consider three kinds of Markovian noise channels on each qubit in
the dynamics, that is, a bit-flip channel, a  phase-flip channel,
and a bit-phase-flip channel. A bit-flip channel can be described
with the two Kraus operators as follows:
\begin{eqnarray}                               
E_1 &=&  \sqrt{1-\frac{p}{2}}\;\; I,
\;\;\;\;\;\;\;\;\;\;\;\;\;\;\;\; E_2 = \sqrt{\frac{p}{2}} \;\;
\sigma_1.\label{eq13}
\end{eqnarray}
The  Kraus operators for a phase-flip channel can be written as
\begin{eqnarray}                               
E_1 &=& \sqrt{1-\frac{p}{2}}\;\; I, \;\;\;\;\;\;\;\;\;\;\;\;\;\;\;\;
E_2 = \sqrt{\frac{p}{2}}\;\; \sigma_3.\label{eq14}
\end{eqnarray}
A bit-phase-flip channel can be described with the following Kraus
operators,
\begin{eqnarray}                               
E_1 &=& \sqrt{1-\frac{p}{2}}\;\; I, \;\;\;\;\;\;\;\;\;\;\;\;\;\;\;\;
E_2 = \sqrt{\frac{p}{2}} \;\; \sigma_2.\label{eq15}
\end{eqnarray}
Here $p$ is the probability that the noise acts on the qubit, and
$0\leq p\leq1$.


The evolved state of a two-qubit system under local environments can
be described with a completely positive trace-preserving map,  which
can be written in the operator-sum representation  \cite{Nielsen}:
\begin{eqnarray}                                
\rho(t)=\sum_{i,j}E^{(A)}_{i}E^{(B)}_{j}\rho(0)E^{(B)^\dagger}_{j}E^{(A)^\dagger}_{i},\label{eq16}
\end{eqnarray}
where $E_i^{(k)}$ $(k=A,B)$  is  the Kraus operator which is used to
describe the channel $A$ or $B$, and $\sum_i
E_i^{(k)}E_i^{(k)^\dagger}=1$.

\subsection*{2.3  $\;\;$   Non-Markovian  noise  channels}

Depolarizing channel is a Markovian noise channel based on bit-flip
error, phase-flip error, and bit-phase-flip error with complete
positive map (CPM)
\begin{eqnarray}                                
\Phi(\rho)=(1-\rho)+\frac{p}{3}(\sigma_1\,\rho\,\sigma_1+\sigma_2\,\rho\,\sigma_2+\sigma_3\,\rho\,\sigma_3),
\end{eqnarray}
where $0\leq p\leq1$. If this noise channel is  generalized to
non-Markovian condition, the memory effects of system-environment
interaction should be introduced. The non-Markovian depolarizing
channel can be derived from generic memory master equation
\begin{eqnarray}                                
\dot{\rho}=K\mathcal {L}\rho,
\end{eqnarray}
where $\mathcal {L}$ is the Lindblad superoperator describing the
dissipative dynamics of system-environment interaction, $K$ is an
integral operator with $K\phi=\int_0^tk(t-t')\phi(t')dt'$, and
kernel function $k(t-t')$ determines the type of memory of
system-environment interaction.

The derivation of memory master equation can be illustrated by the
model with a spin in a magnetic filed \cite{CorrDyna7}. The
time-dependent phenomenological Hamiltonian of this model is
\begin{eqnarray}                               
H(t)=\hbar\,\Gamma_i(t)\,\sigma_i.
\end{eqnarray}
Here $\Gamma_i(t)=a_i\,n_i(t)$ is independent random variable, $a_i$
is a coin-flip random variable taking the values $\pm|a_i|$, and
$n_i$ is a random variable having a Poisson distribution with the
mean value $t/2\tau_i$. With the von Neumann equation
$\dot{\rho}=-(i/\hbar)[H,\rho]$, the memory kernel master equation
is obtained \cite{CorrDyna7},
\begin{eqnarray}                               
\dot{\rho}(t)=-\int_0^t\sum_k
\text{exp}(-(t-t')/\tau_k)\,a_k^2\,[\sigma_k,[\sigma_k,\rho(t')]]dt',
\end{eqnarray}
where
$<\Gamma_j(t)\Gamma_k(t')>=a_k^2exp(-(t-t')/\tau_k)\delta_{jk}$ is
correlation function of the random telegraph signal, $k$ ($=1,2,3$)
represent the general case of noise in three directions.

The linear map $\Phi_t(\rho):\rho\rightarrow\rho_t$ is required to
be a completely positive, trace-preserving map, which has a Kraus
decomposition $\Phi_t(\rho)=\sum_kA_k^\dag\rho A_k$. In
Ref.\cite{CorrDyna7}, this particular condition had been studied in
detail, and the CPM is assured when two of the $a_i$ are zero. The
physical situation of noise is reduced to one direction, and the
Kraus operators of the direction $a_i=a$ are
\begin{eqnarray}                               
A_i=\sqrt{\frac{1-\Lambda(\nu)}{2}}\;\sigma_i,\;\;\;\;\;\;\;\;
A_j=0,\;\;\;\;\;\;\;\; A_k=0,\;\;\;\;\;\;\;\;
A_4=\sqrt{\frac{1+\Lambda(\nu)}{2}}\;I.\label{eq21}
\end{eqnarray}
Here
$\Lambda(t)=\exp(-\upsilon)[\cos(\mu\upsilon)+\sin(\mu\upsilon)/\mu]$
is damped harmonic oscillator, $\mu=\sqrt{(4a\tau)^2-1}$ is
frequency, and $\upsilon=t/(2\tau)$ is dimensionless time. The
non-Markovian depolarizing channel becomes bit-flip channel,
phase-flip channel, and bit-phase-flip channel with $p=1-\Lambda(t)$
in Eqs.(\ref{eq13})-(\ref{eq15}).


\section*{3  $\;\;$  Correlation dynamics of  a two-qubit system in a Bell-diagonal state under  Markovian noise}

\subsection*{3.1  $\;\;$   Bi-side  Markovian noise channels with the same type}

In this section, we discuss the dynamics of correlations of a
two-qubit system which is originally in a Bell-diagonal state and
interacts with two independent local Markovian channels with the
same type but different decoherence rates. There are three types of
noise for local Markovian channels: bit-flip noise, phase-flip
noise, and bit-phase-flip noise.

\bigskip

(1) \emph{Two independent local phase-flip channels}

\bigskip

When two local noise channels are both phase-flip ones, the Kraus
operators  for the two-qubit system are
\begin{eqnarray}                               
E_0^{(A)} &=& \sqrt{1-\frac{p}{2}}\;\; I_A\otimes I_B,
\;\;\;\;\;\;\;\;\;\;\;\;\;\;\;\;\;\;
E_1^{(A)} \;= \; \sqrt{\frac{p}{2}}\;\; \sigma^3_A\otimes I_B,\nonumber\\
E_0^{(B)} &=& I_A\otimes\sqrt{1-\frac{p'}{2}}\;\; I_B,
\;\;\;\;\;\;\;\;\;\;\;\;\;\;\;\;\; E_1^{(B)} \; =\;
I_A\otimes\sqrt{\frac{p'}{2}} \;\; \sigma^3_B,
\end{eqnarray}
where $p=1-\text{exp}(-\gamma t)$, $p'=1-\text{exp}(-\gamma't)$, and
$\gamma$ and $\gamma'$ are the phase damping rates for the channels
$A$ and $B$ (i.e., the qubits $A$ and $B$), respectively. Any
Bell-diagonal state $\rho$ shown in Eq.(\ref{eq7}) evolves to
another Bell-diagonal state under this kind of noise channels,
\begin{eqnarray}                                
\rho(t)=\frac{1}{4}(I+\sum^3_{i=1}c_{i}(t)\sigma_{i}\otimes\sigma_{i}).\label{eq23}
\end{eqnarray}
Here the three coefficients are
\begin{eqnarray}                                
c_1(t) &=& (1-p)(1-p')c_1,\nonumber\\
c_2(t) &=& (1-p)(1-p')c_2,\nonumber\\
c_3(t) &=& c_3.  \label{eq24}
\end{eqnarray}

The dynamics of classical correlation and quantum correlation under
this noise channel are shown in  Eq.(\ref{eq11}) and
Eq.(\ref{eq12}), respectively. Here
$\chi=Max\{|c_1(t)|,|c_2(t)|,|c_3(t)|\}$.

\bigskip

(2) \emph{Two independent local bit-flip channels}

\bigskip

When the two local noise channels are both bit-flip channels, the
Kraus operators of channels for the system are given by
\begin{eqnarray}                               
E_0^{(A)} &=& \sqrt{1-\frac{p}{2}}\;\; I_A\otimes I_B,
\;\;\;\;\;\;\;\;\;\;\;\;\;\;\;\;\;
E_1^{(A)} \;=\; \sqrt{\frac{p}{2}}\;\; \sigma^1_A\otimes I_B,\nonumber\\
E_0^{(B)} &=& I_A\otimes\sqrt{1-\frac{p'}{2}} \;\; I_B,
\;\;\;\;\;\;\;\;\;\;\;\;\;\;\;\; E_1^{(B)} \;=\;
I_A\otimes\sqrt{\frac{p'}{2}} \;\;\sigma^1_B.
\end{eqnarray}
If the initial state is Bell-diagonal state (Eq.(\ref{eq7})), under
this noise channel, the coefficients may decay to be
\begin{eqnarray}                                
c_1(t) &=& c_1,\nonumber\\
c_2(t) &=& (1-p)(1-p')c_2,\nonumber\\
c_3(t) &=& (1-p)(1-p')c_3. \label{eq26}
\end{eqnarray}

\bigskip

(3)  \emph{Two independent local bit-phase-flip channels}

\bigskip

When the two local noise channels are both bit-phase-flip channels,
the Kraus operators are
\begin{eqnarray}                               
E_0^{(A)} &=& \sqrt{1-\frac{p}{2}}\;\; I_A\otimes I_B,
\;\;\;\;\;\;\;\;\;\;\;\;\;\;\;\;\;
E_1^{(A)} \;=\;  \sqrt{\frac{p}{2}} \;\; \sigma^2_A\otimes I_B,\nonumber\\
E_0^{(B)} &=& I_A\otimes\sqrt{1-\frac{p'}{2}} \;\;  I_B,
\;\;\;\;\;\;\;\;\;\;\;\;\;\;\;\; E_1^{(B)} \;=\;
I_A\otimes\sqrt{\frac{p'}{2}} \;\;  \sigma^2_B.
\end{eqnarray}
The initial state shown in Eq.(\ref{eq7}) under these noise channels
can be evolved to another Bell-diagonal state (Eq.(\ref{eq23})) with
the coefficients
\begin{eqnarray}                                
c_1(t) &=& (1-p)(1-p')c_1,\nonumber\\
c_2(t) &=& c_2,\nonumber\\
c_3(t) &=& (1-p)(1-p')c_3. \label{eq28}
\end{eqnarray}

\begin{figure}                                    
\begin{minipage}[t]{0.33\linewidth}
\centering
\includegraphics[width=2.2 in]{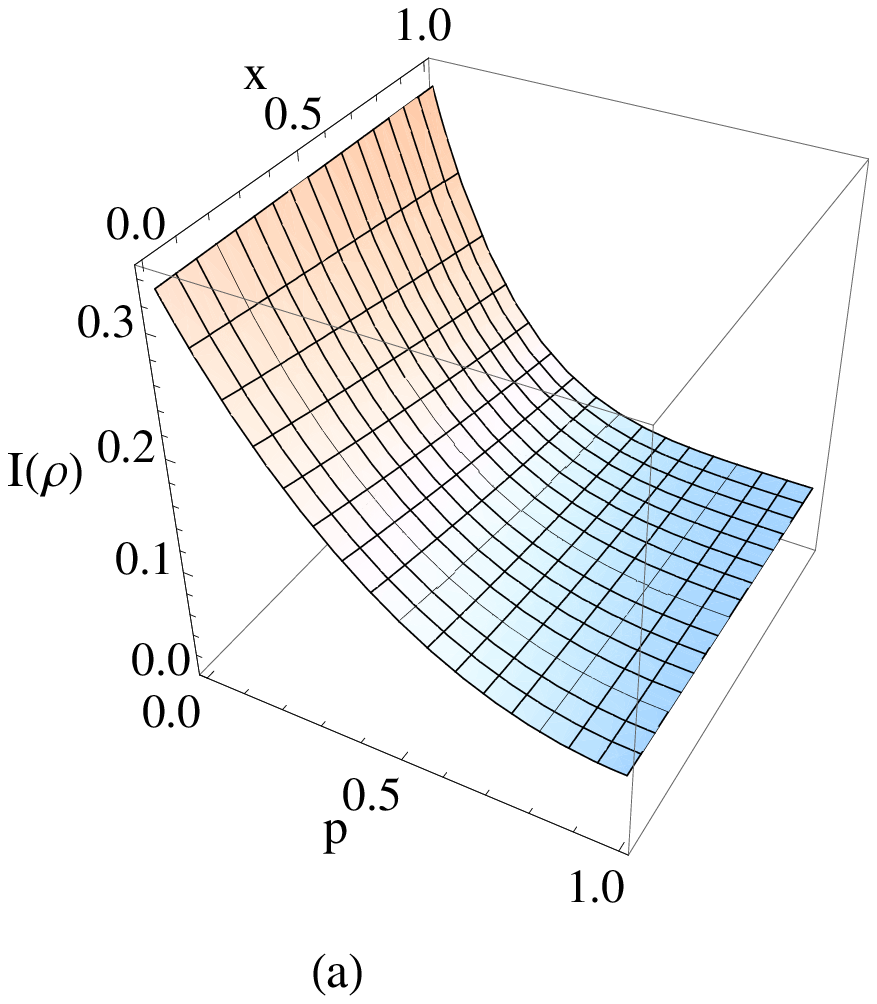}
\end{minipage}%
\begin{minipage}[t]{0.33\linewidth}
\centering
\includegraphics[width=2.2 in]{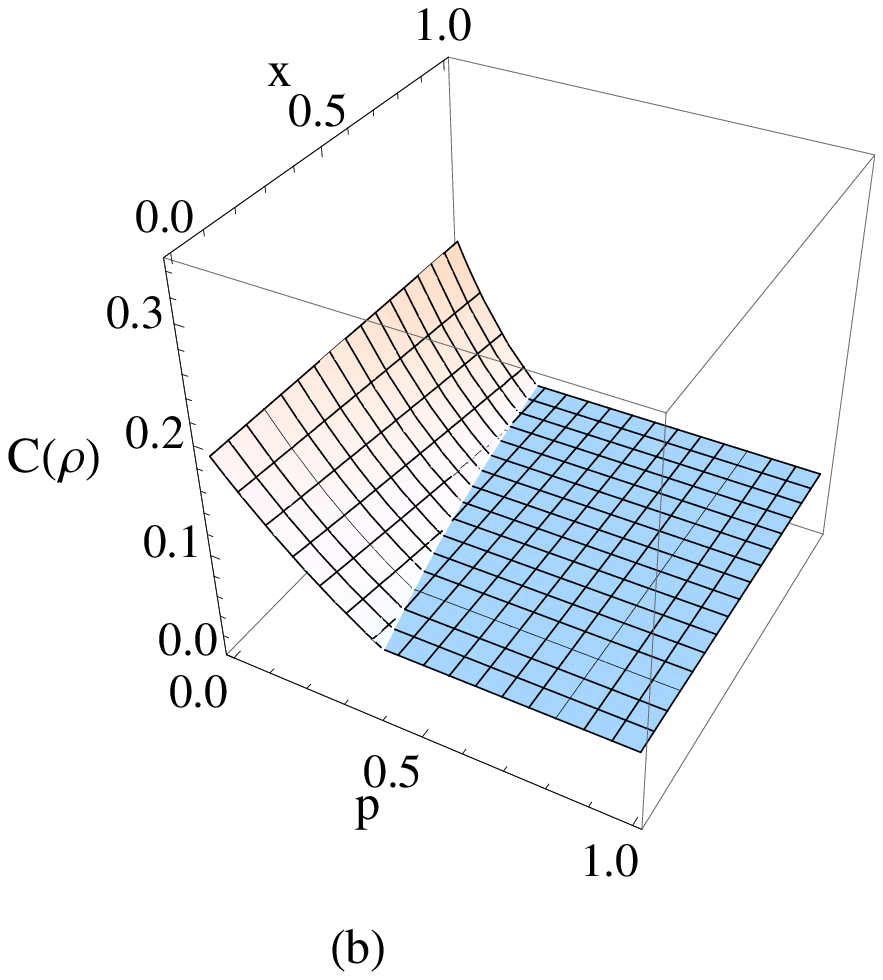}
\end{minipage}
\begin{minipage}[t]{0.33\linewidth}
\centering
\includegraphics[width=2.2 in]{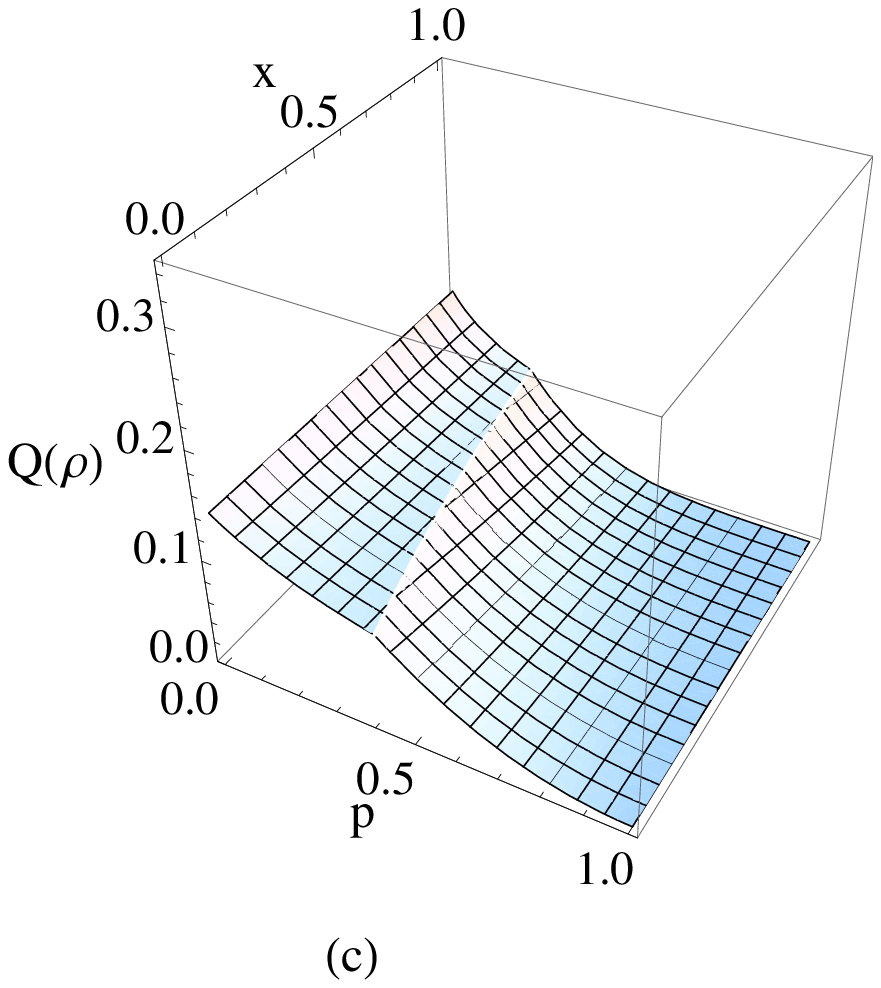}
\end{minipage}
\caption{(Color online) The dynamics of mutual information (a),
classical correlation (b), and quantum correlation (c) of a
two-qubit system in the state $\rho$ (Eq.(\ref{eq23})) with initial
coefficients $c_1=0.1$, $c_2=0.5$, $c_3=0.3$ under the  bi-side
phase-flip channels. Here $x=p'/p$ represents the relation of two
parameter times of two phase-flip channels. }\label{fig1}
\end{figure}


The three noise channels discussed above evolve one Bell-diagonal
state to another Bell-diagonal state by decaying coefficients
$\{c_i\}$. From Eq.(\ref{eq11}) and (\ref{eq12}), one can see that
the classical correlation is dependent on the maximal value of
coefficients $\{|c_i|\}$, and the quantum correlation is the
difference of mutual information and classical correlation. The
decay rate of classical correlation  depends on the decay rate of
maximal coefficients $\{|c_i(t)|\}$. For example, if the noise of
local channels are both phase-flip noise, the dynamics of classical
and quantum correlations have three conditions as in
Ref.\cite{CorrDyna1}.

(a)  When $|c_3|=0$, the quantum correlation and classical
correlation decay monotonically.

(b) When $|c_3|\ge\{|c_1|,|c_2|\}$, the classical correlation will
not change in this process, and quantum correlation has the same
decay rate as mutual information.

\begin{figure}                                    
\begin{minipage}[t]{0.33\linewidth}
\centering
\includegraphics[width=2.2 in]{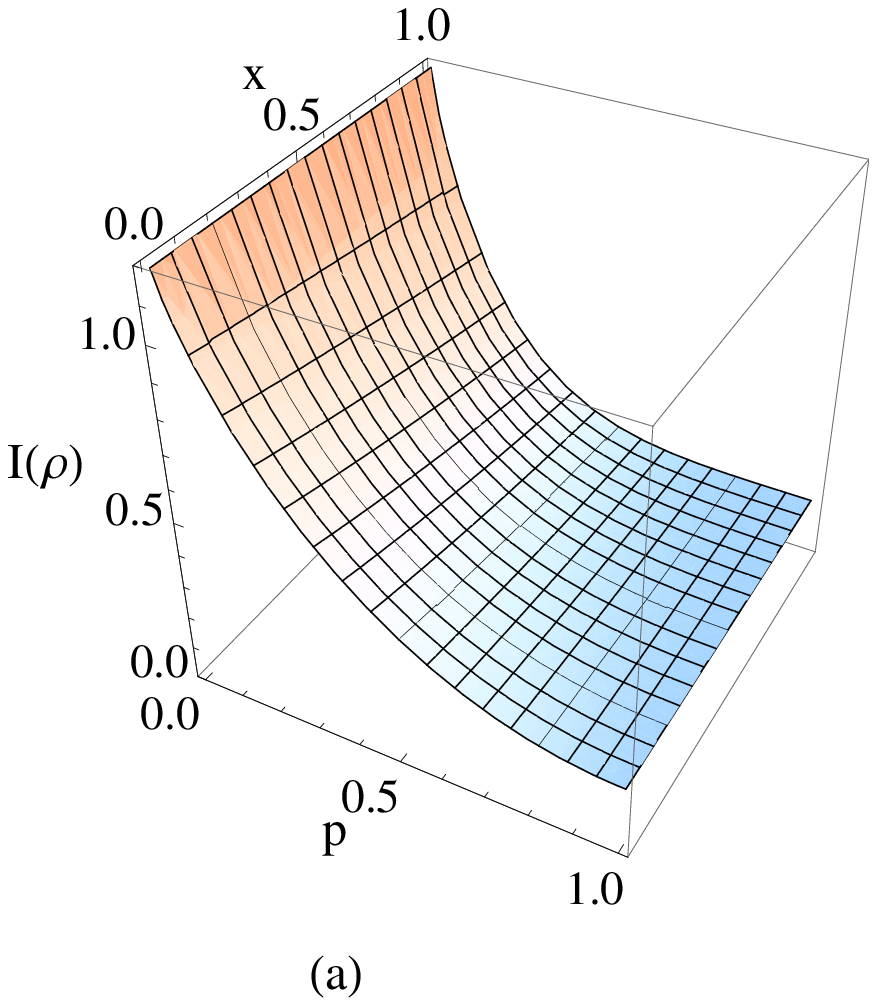}
\end{minipage}%
\begin{minipage}[t]{0.33\linewidth}
\centering
\includegraphics[width=2.2 in]{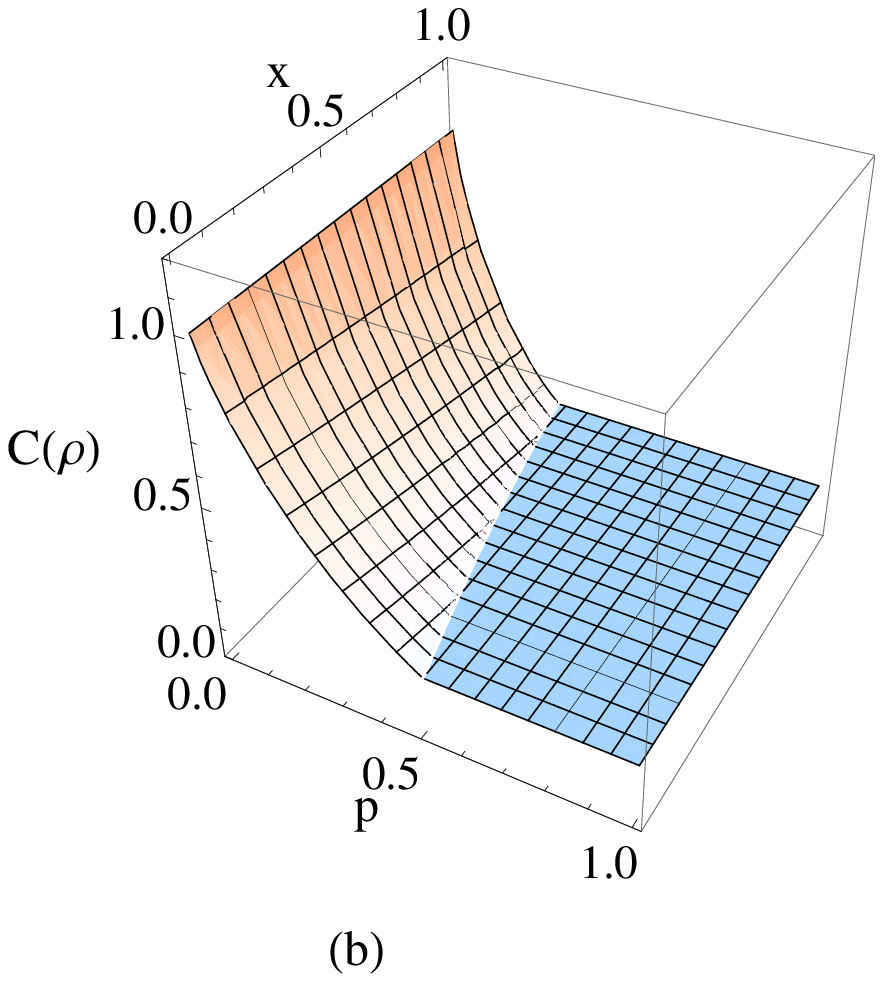}
\end{minipage}
\begin{minipage}[t]{0.33\linewidth}
\centering
\includegraphics[width=2.2 in]{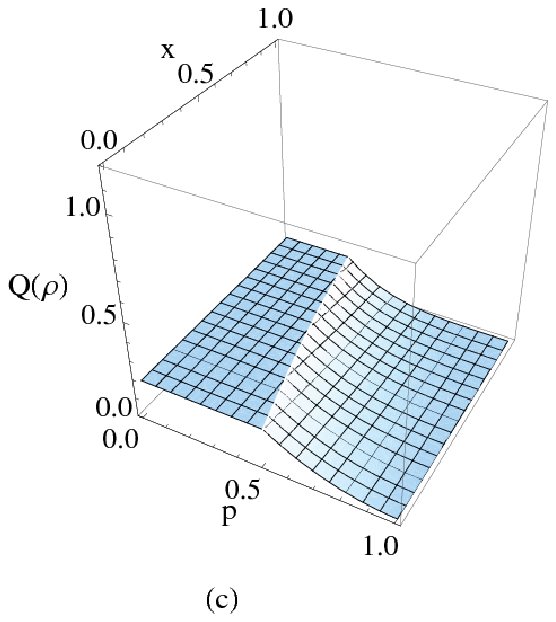}
\end{minipage}
\caption{(Color online) The dynamics of mutual information (a),
classical correlation (b), and quantum correlation (c) of a
two-qubit system in the state $\rho(t)$ (Eq.(\ref{eq23})) with
initial coefficients $c_1=1$, $c_2=-0.5$, $c_3=0.5$ under the
bi-side phase flip channels. Here $x=p'/p$ represents the relation
of two parameter times of two phase-flip channels.}\label{fig2}
\end{figure}

(c)  When $|c_2|\ge\{|c_1|,|c_3|\}$ ($|c_1|\ge\{|c_2|,|c_3|\}$) and
$|c_3|\neq0$, the classical correlation decays with the diminishing
of $c_2(t)$ ($c_1(t)$) before the point $c_3=(1-p)(1-p')c_2$
($c_3=(1-p)(1-p')c_1$), and then classical correlation suddenly
changes to a constant. While the quantum correlation in this
condition has the decay rate which is suddenly changed at the same
point $c_3=(1-p)(1-p')c_2$ ($c_3=(1-p)(1-p')c_1$). Figure \ref{fig1}
shows the dynamics of mutual information, quantum correlation and
classical correlation for the symmetry phase-flip channels. The
relation of decay rates of the two local channels is $p'=xp$. The
decay rates of quantum and classical correlations become larger and
the sudden change point is moved forward with $x$ changing from $0$
to $1$. For some special cases $c_1=k$, $c_2=-c_3k$ and $|k|>|c_3|$
\cite{CorrDyna8}, the sudden transition between classical and
quantum decoherence takes place, as shown in Fig.\ref{fig2}.

The three symmetry noise channels with phase-flip, bit-flip and
bit-phase-flip are equivalent to local unitary operations.
Therefore, the dynamics of classical and quantum correlations under
the other two channels with the time parameters $p$ and $p'$ are
similar to those in the case with symmetry phase-flip channels. For
bit-flip channels, the relation of coefficients for three regimes
are $|c_1|=0$, $|c_1|\ge\{|c_2|,|c_3|\}$ and
$|c_2|\ge\{|c_1|,|c_3|\}$ ($|c_3|\ge\{|c_1|,|c_2|\}$). For
bit-phase-flip channels, the relation of coefficients for three
regimes are $|c_2|=0$, $|c_2|\ge\{|c_1|,|c_3|\}$ and
$|c_1|\ge\{|c_2|,|c_3|\}$ ($|c_3|\ge\{|c_1|,|c_2|\}$).


\subsection*{3.2  $\;\;$   Two local channels are different types}

In this section, we discuss the dynamics of correlations of a
two-qubit system in  a Bell-diagonal state under two different local
Markovian noise channels. That is, the noise channels interacting
with the two qubits are different types and have different
decoherence rates. The local Markovian noise channels discussed here
are still phase-flip channel, bit-flip channel, and bit-phase-flip
channel, respectively.

\bigskip

(1) \emph{A bit-flip channel and a phase-flip channel}

\bigskip

When the two local Markovian noise channels  are a bit-flip channel
and a phase-flip channel, the Kraus operators for two-qubit sysem
can be written as
\begin{eqnarray}                               
E_0^{(A)} &=& \sqrt{1-\frac{p}{2}}\;\; I_A\otimes I_B,
\;\;\;\;\;\;\;\;\;\;\;\;\;\;\;\;\;
E_1^{(A)} \;=\; \sqrt{\frac{p}{2}}\;\;  \sigma^1_A\otimes I_B,\nonumber\\
E_0^{(B)} &=& I_A\otimes\sqrt{1-\frac{p'}{2}}\;\;  I_B,
\;\;\;\;\;\;\;\;\;\;\;\;\;\;\;\; E_1^{(B)}
\;=\;I_A\otimes\sqrt{\frac{p'}{2}}\;\; \sigma^3_B.
\end{eqnarray}
The coefficients of the final two-qubit Bell-diagonal state are
\begin{eqnarray}                              
c_1(t) &=& (1-p')c_1,\nonumber\\
c_2(t) &=& (1-p)(1-p')c_2,\nonumber\\
c_3(t) &=& (1-p)c_3.\label{eq30}
\end{eqnarray}

\bigskip

(2) \emph{A bit-flip channel and a bit-phase-flip channel}

\bigskip

When the two local channels are a bit-flip one and a bit-phase-flip
one, the Kraus operators for these channels are
\begin{eqnarray}                               
E_0^{(A)} &=& \sqrt{1-\frac{p}{2}}\;\; I_A\otimes I_B,
\;\;\;\;\;\;\;\;\;\;\;\;\;\;\;\;\;
E_1^{(A)} \;=\; \sqrt{\frac{p}{2}}\;\;  \sigma^1_A\otimes I_B,\nonumber\\
E_0^{(B)} &=& I_A\otimes\sqrt{1-\frac{p'}{2}} \;\;  I_B,
\;\;\;\;\;\;\;\;\;\;\;\;\;\;\;\; E_1^{(B)} \;=\;
I_A\otimes\sqrt{\frac{p'}{2}}\;\; \sigma^2_B.
\end{eqnarray}
The coefficients of the final two-qubit Bell-diagonal state decay to
be
\begin{eqnarray}                              
c_1(t) &=& (1-p')c_1,\nonumber\\
c_2(t) &=& (1-p)c_2,\nonumber\\
c_3(t) &=& (1-p)(1-p')c_3. \label{eq32}
\end{eqnarray}

\bigskip

(3) \emph{A phase-flip channel and a bit-phase-flip channel}

\bigskip

When the two local channels are a phase-flip channel and a
bit-phase-flip channel, the Kraus operators of these channels are
\begin{eqnarray}                               
E_0^{(A)} &=& \sqrt{1-\frac{p}{2}}\;\; I_A\otimes I_B,
\;\;\;\;\;\;\;\;\;\;\;\;\;\;\;\;\;
E_1^{(A)} \;=\; \sqrt{\frac{p}{2}}\;\; \sigma^3_A\otimes I_B,\nonumber\\
E_0^{(B)} &=& I_A\otimes\sqrt{1-\frac{p'}{2}} \;\; I_B,
\;\;\;\;\;\;\;\;\;\;\;\;\;\;\;\; E_1^{(B)} \;=\;
I_A\otimes\sqrt{\frac{p'}{2}}\;\; \sigma^2_B.
\end{eqnarray}
The coefficients of the final two-qubit Bell-diagonal state are
\begin{eqnarray}                               
c_1(t) &=& (1-p)(1-p')c_1,\nonumber\\
c_2(t) &=& (1-p)c_2, \nonumber\\
c_3(t) &=& (1-p')c_3. \label{eq34}
\end{eqnarray}

The two-side noise channels discussed above have different types and
decoherence rates for the two qubits in the system.  These three
kinds of noise channels  evolve one Bell-diagonal state to another
Bell-diagonal state and change the coefficients $\{c_i\}$ yet.
Different from the cases under the noises with the same types, the
correlations are not symmetry for the parameter times $p$ and $p'$.
Here the relation of parameter times of two local channels is
$p'=xp$ and $0\leq x \leq1$.

\begin{figure}                                    
\begin{minipage}[t]{0.33\linewidth}
\centering
\includegraphics[width=2.2 in]{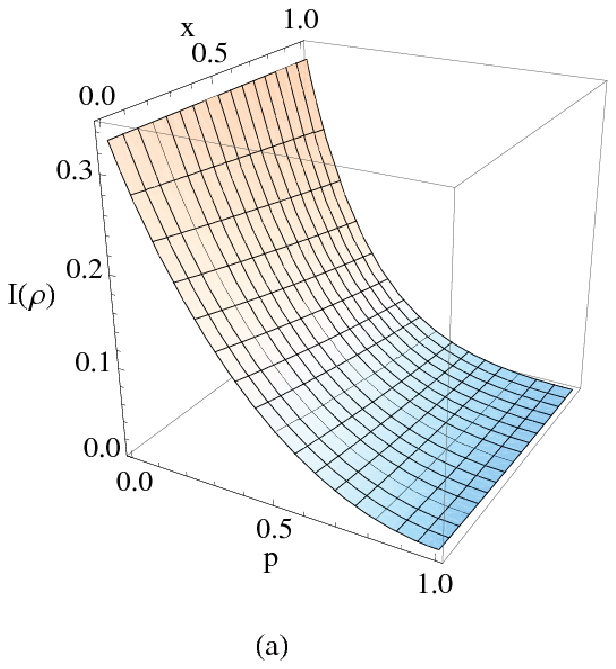}
\end{minipage}%
\begin{minipage}[t]{0.33\linewidth}
\centering
\includegraphics[width=2.2 in]{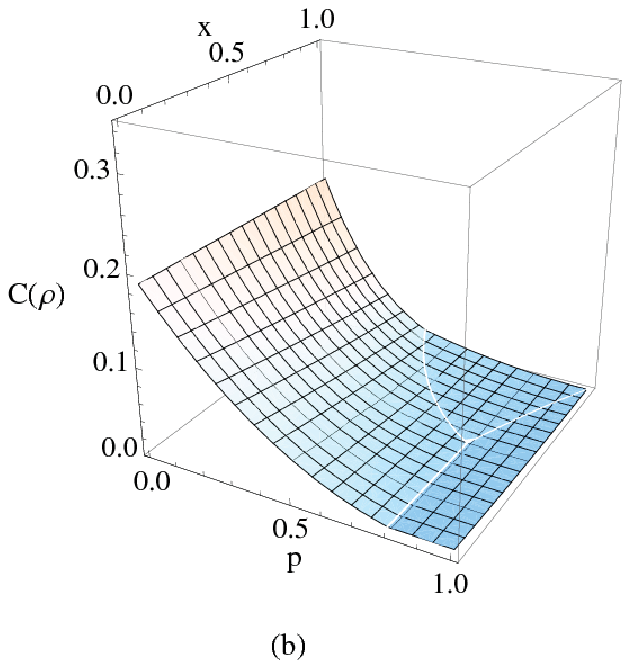}
\end{minipage}
\begin{minipage}[t]{0.33\linewidth}
\centering
\includegraphics[width=2.2 in]{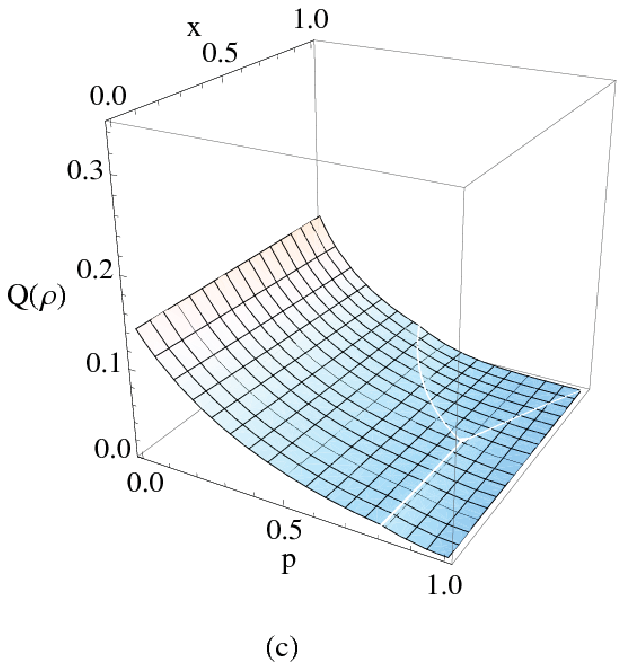}
\end{minipage}
\caption{(Color online) The dynamics of mutual information (a),
classical correlation (b), and quantum correlation (c) of a
two-qubit system in the state $\rho(t)$ with the parameters
$c_1=0.1$, $c_2=0.5$, $c_3=0.3$ under the bi-side Markovian noise
composed of a bit-flip channel and a phase-flip channel. Here
$x=p'/p$ represents the relation of two parameter times of two local
noise channels.}\label{fig3}
\end{figure}

If the noise channels of the two sides are a bit-flip channel and a
phase-flip channel, there are still three conditions with the
different relation of coefficients $\{c_i\}$.

(a) When $|c_1|\ge\{|c_2|,|c_3|\}$, the relation of the coefficients
$\{c_i(t)\}$ are $|c_1(t)|\ge\{|c_2(t)|,|c_3(t)|\}$, and both the
quantum correlation and classical correlation decay monotonically.

(b) When $|c_3|\ge\{|c_1|,|c_2|\}$ ($|c_2|\ge|c_1|\ge|c_3|$), the
decay rates of the quantum correlation and classical correlation are
both suddenly changed at the point $c_1(1-p')=(1-p)c_3$
($c_2(1-p)=c_1$). In the case $|c_3|\ge\{|c_1|,|c_2|\}$, the sudden
change point is moved backward and the decay rates of quantum
correlation and classical correlation become larger with $x$
changing from $0$ to $1$. In the case $|c_2|\ge|c_1|\geq|c_3|$, the
sudden change point is not affected by $x$, while the decay rates of
quantum correlation and classical correlation still become larger
with $x$ changing from $0$ to $1$.

(c) When $|c_2|\ge|c_3|\geq|c_1|$, the sudden change points are
determined by $x$. If $x\leq(c_2-c_3)/(c_2-c_1)$, the decay rates of
quantum correlation and classical correlation are suddenly changed
at the point $c_2(1-p)=c_1$. This sudden change point is not
affected by $x$, while the decay rates of quantum correlation and
classical correlation become larger with $x$ changing from $0$ to
$(c_2-c_3)/(c_2-c_1)$. If $x\geq(c_2-c_3)/(c_2-c_1)$, the decay
rates of quantum correlation and classical correlation have two
sudden change points at $c_2(1-p')=c_3$ and $c_3(1-p)=c_1(1-p')$.
The sudden change  point $c_2(1-p')=c_3$ is moved forward and the
decay rates of quantum correlation and classical correlation become
larger with $x$ changing from $(c_2-c_3)/(c_2-c_1)$ to $1$. The
sudden change point $c_3(1-p)=c_1(1-p')$ is moved backward and the
decay rates of quantum correlation and classical correlation become
larger with $x$ changing from $(c_2-c_3)/(c_2-c_1)$ to $1$.

The three noise channels with phase-flip, bit-flip, and
bit-phase-flip are equivalent to local unitary operations.
Therefore, the dynamics of classical and quantum correlations under
the other two channels with parameter times $p$ and $p'$ are similar
to the noise channels with a bit-flip channel and a phase-flip
channel. For the condition of a bit-flip and a bit-phase-flip
channels, the relation of coefficients for three regimes are
$|c_1|\ge\{|c_2|,|c_3|\}$, $|c_2|\ge\{|c_1|,|c_3|\}$
($|c_3|\ge|c_1|\ge|c_2|$), and $|c_3|\ge|c_2|\geq|c_1|$. For the
condition of a phase-flip and a bit-phase-flip channels, the
relation of coefficients for three regimes are
$|c_3|\ge\{|c_1|,|c_2|\}$, $|c_2|\ge\{|c_1|,|c_3|\}$
($|c_1|\ge|c_3|\ge|c_2|$), and $|c_1|\ge|c_2|\geq|c_3|$.


\section*{4  $\;\;$  Correlation dynamics of  a two-qubit system in a Bell-diagonal state under non-Markovian
noise}

As introduced in Sec. 2.3, the Markovian channel with phase-flip
noise, bit-flip noise, and bit-phase-flip noise can be generalized
to non-Markovian channel with parameter time oscillating as
$p=1-\Lambda(t)$. Due to the Kraus composition form of non-Markovian
channels, the finial state after operating on Bell-diagonal state
satisfies Eq.(\ref{eq23}). Here we still investigate the noise
channels of two sides in two conditions, bi-side  non-Markovian
noise channels with the same type or different types.

\begin{figure}                                    
\begin{minipage}[t]{0.33\linewidth}
\centering
\includegraphics[width=2.2 in]{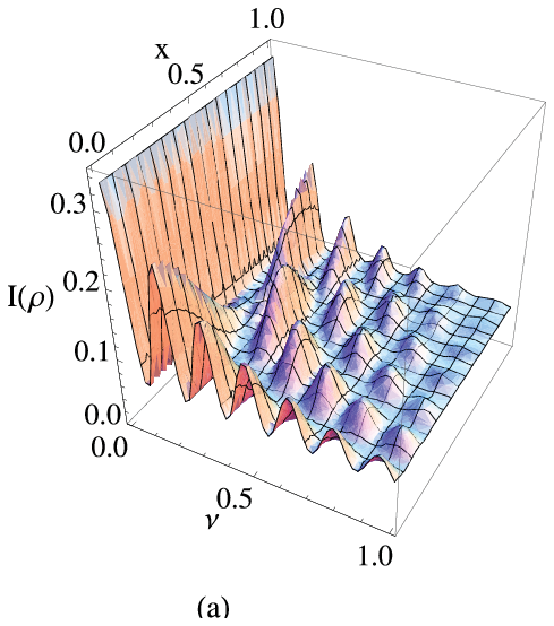}
\end{minipage}%
\begin{minipage}[t]{0.33\linewidth}
\centering
\includegraphics[width=2.2 in]{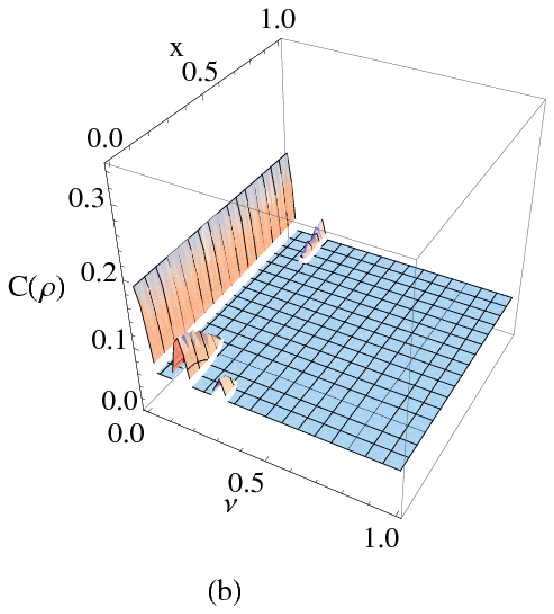}
\end{minipage}
\begin{minipage}[t]{0.33\linewidth}
\centering
\includegraphics[width=2.2 in]{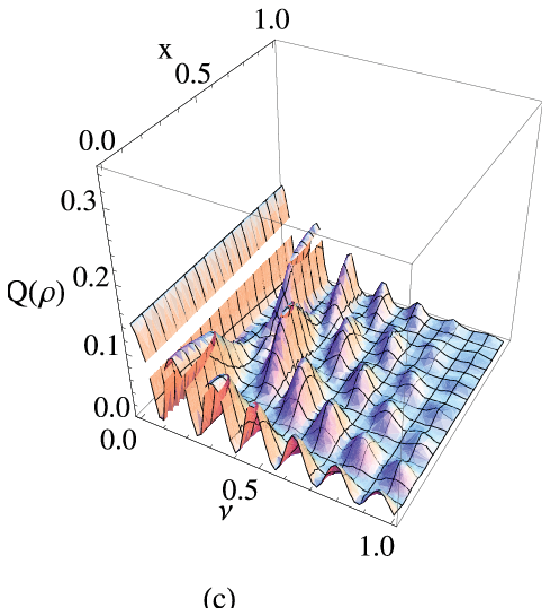}
\end{minipage}
\caption{(Color online) Dynamics of mutual information (a),
classical correlation (b), and quantum correlation (c) of two-qubit
system under noise channel of which two sides are both non-Markovian
phase-flip channels. The initial coefficients of two-qubit system
are $c_1=0.1$, $c_2=0.5$, $c_3=0.3$, and the parameters of
non-Markovian phase-flip channel are $\tau=5s$, $a=1s^{-1}$. Here
$x=\nu'/\nu$ represents the relation of two parameter times of two
non-Markovian phase-flip channels.}\label{fig4}
\end{figure}

\bigskip

(1) \emph{Bi-side non-Markovian noise channels with the same type}

\bigskip

Here the three local non-Markovian channel  are bit-flip,
phase-flip, and bit-phase-flip channels oscillating with time
parameter $\nu$. The non-Markovian noise channels of two sides are
the same type with different time parameter $\nu'=x\nu$. As
discussed in Sec. 3.1, as the three types of noise channels are
equivalent to each other, we focus on the condition of bi-side
non-Markovian phase-flip channels.

Using the Kraus operators in Eq.(\ref{eq21}), the finial two-qubit
Bell-diagonal state in Eq.(\ref{eq23}) can be calculated with
coefficients evolving as
\begin{eqnarray}                                
c_1(t) &=& \Lambda(\nu)\Lambda(\nu')c_1,\nonumber\\
c_2(t) &=& \Lambda(\nu)\Lambda(\nu')c_2,\nonumber\\
c_3(t) &=& c_3.  \label{eq35}
\end{eqnarray}
The dynamics of correlations can still be distinguished to three
regimes, as the same as Markovian channels corresponding to the
relation of coefficients $\{c_i\}$.

(a) $|c_3|=0$. The mutual information, quantum correlation and
classical correlation display damped oscillations and decay
asymptotically to zero.

(b) $|c_3|\ge\{|c_1|,|c_2|\}$. The classical correlation will not
change in this regime, and quantum correlation has the same damped
oscillations with mutual information.

\begin{figure}                                    
\begin{minipage}[t]{0.33\linewidth}
\centering
\includegraphics[width=2.2 in]{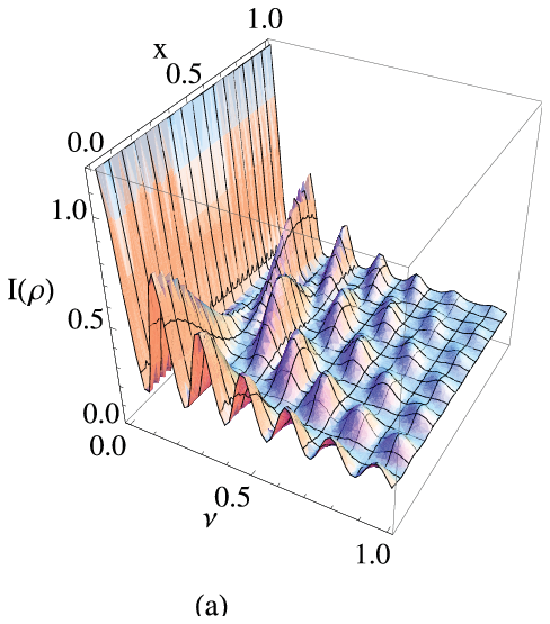}
\end{minipage}%
\begin{minipage}[t]{0.33\linewidth}
\centering
\includegraphics[width=2.2 in]{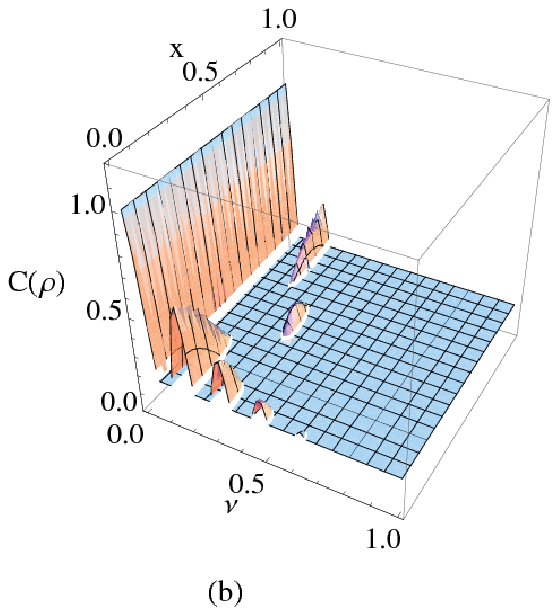}
\end{minipage}
\begin{minipage}[t]{0.33\linewidth}
\centering
\includegraphics[width=2.2 in]{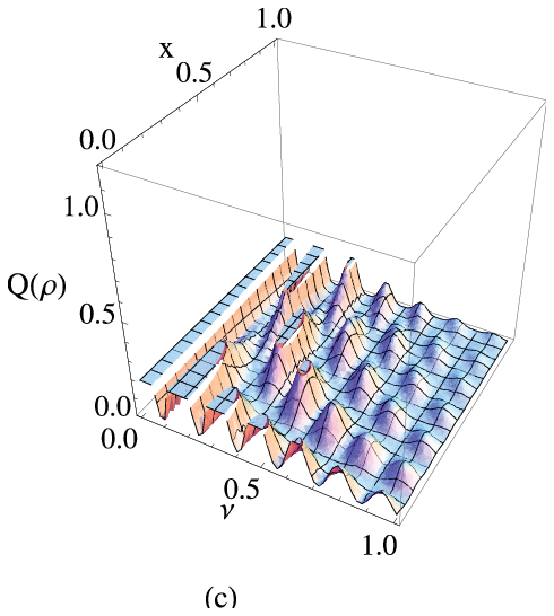}
\end{minipage}
\caption{(Color online) Dynamics of mutual information (a),
classical correlation (b), and quantum correlation (c) of two-qubit
system under noise channel of which two sides are both non-Markovian
phase-flip channels. The initial coefficients of two-qubit system
are $c_1=1$, $c_2=-0.5$, $c_3=0.5$, and the parameters of
non-Markovian phase-flip channel are $\tau=5s$, $a=1s^{-1}$. Here
$x=\nu'/\nu$ represents the relation of two parameter times of two
non-Markovian phase-flip channels.}\label{fig5}
\end{figure}

(c) $|c_2|\ge\{|c_1|,|c_3|\}$ ($|c_1|\ge\{|c_2|,|c_3|\}$) and
$|c_3|\neq0$. The classical correlation decays with damped
oscillation of $c_2(t)$ ($c_1(t)$) and changes to be a constant at
sudden change points $c_3=\Lambda(\nu)\Lambda(\nu')c_2$
($c_3=\Lambda(\nu)\Lambda(\nu')c_1$). The quantum correlation has
decay rate changed  suddenly at these sudden change points. As shown
in Fig.\ref{fig4}, the classical and quantum correlations decrease
to the sudden change point, and then classical correlation becomes
constant while quantum correlation has a discontinuous change with
the changing of damping oscillation rule. There is a special case
with $c_1=k$, $c_2=-c_3k$ and $|k|>|c_3|$ as shown in
Fig.\ref{fig5}, the quantum correlation may become constant at a
sudden change point which is called frozen discord in Ref.
\cite{CorrDyna8}. Corresponding to the sudden transition between
classical and quantum decoherence of Markovian channels, the
transition between classical and quantum damping oscillations
happens.

\bigskip

(2) \emph{Bi-side non-Markovian noise channels with different types}

\bigskip

Now we discuss the dynamics of correlations of two local
non-Markovian noise channels with different types operating on a
two-qubit Bell-diagonal state. The three types of non-Markovian
channels are bit-flip, phase-flip, and bit-phase-flip channels
oscillating with parameter time $\nu$. We focus on two-side noise
channels composed of a bit-flip and a phase-flip channel with
different parameter time $\nu'=x\nu$.

The finial Bell-diagonal state after this noise channel is
\begin{eqnarray}                           
c_1(t) &=& \Lambda(\nu')c_1,\nonumber\\
c_2(t) &=& \Lambda(\nu)\Lambda(\nu')c_2,\nonumber\\
c_3(t) &=& \Lambda(\nu)c_3.\label{eq36}
\end{eqnarray}
The dynamics of correlations is determined by the coefficients
$\{c_i\}$ as we discussed above. If $x=0$, the parameter time of a
phase-flip channel is $\Lambda(0)=1$, and the noise channel becomes
a one-side noise channel. For this condition, the dynamics of
correlations in three regimes are the same as bi-side non-Markovian
noise channels with the same type. If $x\neq0$, the oscillation
periods of $\Lambda(\nu)$ and $\Lambda(\nu')$ are decided by the
parameter times of the two local channels $\nu$ and $\nu'$. The
different oscillation periods of $\Lambda(\nu)$ and $\Lambda(\nu')$
implies the uncontinuous changes of both classical and quantum
correlations may happen in any relation regime of coefficients
$\{c_i\}$.


\section*{5 $\;\;$  Discussion and conclusion}

In this article, we have studied the dynamics of correlations of a
two-qubit system in a Bell-diagonal state under independent local
noise channels of which the two sides are not identical. With the
anti-commute relation of Pauli operators, the Bell-diagonal state
has the coefficients $\{c_i\}$ decreased when the bi-side local
noise channels are phase-flip channel, bit-flip channel, and
bit-phase-flip channel, respectively. The classical correlation is
determined by the maximal value of $\{|c_i(t)|\}$.

Under Markovian noise channels, the dynamics of correlations
decrease monotonically. If the two sides of noise channels are the
same type, there are three regimes for the decay rates, dependent on
the coefficients $\{c_i\}$. In the first regime, both the classical
and quantum correlations decay monotonically asymptotically to zero.
In the second regime, the classical correlation keeps to be a
constant, while the quantum correlation decreases with the same
decay rate as mutual information. In the third regime, classical
correlation decreases monotonically till the sudden change point
changing to a constant, while the quantum correlation has the decay
rate changed suddenly at the sudden change point. If the two sides
of noise channels are the different types, there are also three
regimes for the decay rates with relation of coefficients $\{c_i\}$.
In the first regime, both the classical and quantum correlations
decay monotonically. In the second regime, both the classical and
quantum correlations have decay rates sudden changed at the sudden
change point. In the third regime, there appears the condition of
two sudden change points. The sudden change point is affected by the
parameter $x$, which describes the relation of two time parameters
$p$ and $p'$, and the regime of $x$ for two sudden change points is
calculated.

When the two sides are non-Markovian channels, correlations may
display damped oscillations. If the two sides of noise channels are
the same type, there are still three regimes for the decay rates,
dependent on the coefficients $\{c_i\}$. In the first regime, both
the quantum correlation and classical correlation display damped
oscillations and decay asymptotically to zero. In the second regime,
the classical correlation keeps a constant while the quantum
correlation has the same damped oscillations with mutual
information. In the third regime, classical correlation may be a
constant in some time intervals between sudden change points, while
quantum correlation has uncontinuous changes with the changing of
damping oscillation rule at the sudden change points. If the two
sides of noise channels are the different types, both classical and
quantum correlations have uncontinuous  changes with the changing of
damping oscillation rule at the sudden change points when the
parameter times of the two local channels are different. The sudden
change points are affected by parameter $x$ which describes the
relation of  time parameters $\nu$ and $\nu'$.


\section*{Acknowledgments}

This work was supported by the National Natural Science Foundation
of China under Grant No. 11174039, NCET-11-0031, and the Fundamental
Research Funds for the Central Universities.

\end{document}